\documentclass{article}
\usepackage{graphicx}
\usepackage{dcolumn}
\usepackage{bm}
\begin{document}
\title{\Large Analyses of $dN_{ch}/d\eta$ and $dN_{ch}/dy$ distributions of BRAHMS Collaboration by means of the Ornstein-Uhlenbeck process
}

\author{\normalsize M. Ide$^1$, M. Biyajima$^{1,2}$ and T. Mizoguchi$^3$\\
$^1$Department of Physics, Faculty of Science, Shinshu University, \\
Matsumoto 390-8621, Japan\\
$^2$The Niels Bohr Institute, DK-2100, Copenhagen, Denmark\\
$^3$Toba National College of Maritime Technology, Toba 517-8501, Japan}

\date{\today}
\maketitle

\begin{abstract}
Interesting data on $dN_{\rm ch}/d\eta$ in Au-Au collisions ($\eta=-\ln \tan (\theta/2)$) with the centrality cuts have been reported by  BRAHMS Collaboration.
Using the total multiplicity $N_{\rm ch} = \int (dN_{\rm ch}/d\eta)d\eta$, we find that there are scaling phenomena among $(N_{\rm ch})^{-1}dN_{\rm ch}/d\eta = dn/d\eta$ with different centrality cuts at $\sqrt{s_{NN}} =$ 130 GeV and 200 GeV, respectively. To explain these scaling behaviors of $dn/d\eta$, we consider the stochastic approach named the Ornstein-Uhlenbeck process with two sources. The following Fokker-Planck equation is adopted for the present analyses,
$$
\frac{\partial P(x,t)}{\partial t} = \gamma \left[\frac{\partial}{\partial x}x + \frac 12\frac{\sigma^2}{\gamma}\frac{\partial^2}{\partial x^2}\right] P(x,\, t)
$$
where $x$ means the rapidity (y) or pseudo-rapidity ($\eta$). $t$, $\gamma$ and $\sigma^2$ are the evolution parameter, the frictional coefficient and the variance, respectively.
Introducing a variable of $z_r = \eta/\eta_{\rm rms}$ ($\eta_{\rm rms}=\sqrt{\langle \eta^2 \rangle}$) we explain the $dn/d z_r$ distributions in the present approach. Moreover, to explain the rapidity (y) distributions from $\eta$ distributions at 200 GeV, we have derived the formula as 
$$
\frac{dn}{dy}=J^{-1}\frac{dn}{d \eta} ,
$$
where $J^{-1}=\sqrt{M(1+\sinh^2 y)}/\sqrt{1+M\sinh^2 y}$ with $M = 1 + (m/p_{\rm t})^2$. Their data of pion and all hadrons are fairly well explained by the O-U process. To compare our approach with another one, a phenomenological formula by Eskola et al. is also used in calculations of $dn/d\eta$.
\end{abstract}

%
\section{Introduction}
Recently interesting data on $dN_{\rm ch}/d\eta$ ($\eta=-\ln \tan (\theta/2)$) and $(0.5\langle N_{\rm part}\rangle)^{-1}dN_{\rm ch}/d\eta|_{\eta=0}$ in Au+Au collision at $\sqrt{s_{NN}} = 130$ GeV and 200 GeV have been reported by BRAHMS Collaboration \cite{Bearden:2001xw,Bearden:2001qq}. ($\langle N_{\rm part}\rangle$ and $N_{\rm ch}$ mean the numbers of participants (nuclei) and charged particles produced in collisions, respectively.) Very recently the BRAHMS Collaboration has reported preliminary data on rapidity (y) distribution at 200 GeV in Ref.~\cite{Ouerdane:2002gm}. We are interested in theoretical analyses of these data.

On the other hand, in Refs.~\cite{Biyajima:2002at,Biyajima:2002ab} we have investigated the property of $\eta$ scaling of $(N_{\rm ch})^{-1}dN_{\rm ch}/d\eta = dn/d\eta$ by PHOBOS Collaboration and found that the $\eta$ scaling holds. As a possible theoretical approach, we have adopted the stochastic theory named the Ornstein-Uhlenbeck (O-U) process with two sources at $\pm y_{\rm max}=\ln(\sqrt{s_{NN}}/m_{N})$. In this paper, we would like to analyses data \cite{Bearden:2001xw,Bearden:2001qq,Ouerdane:2002gm} by the stochastic approach in terms of the pseudo-rapidity and/or rapidity variables.

\begin{figure}
\begin{center}
  \includegraphics[height=40mm]{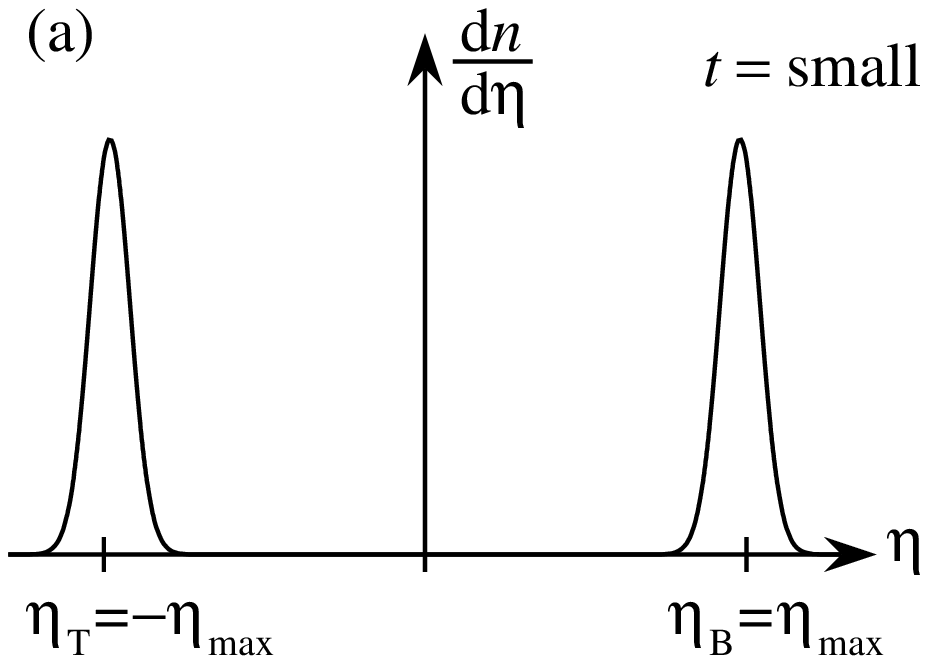}
  \includegraphics[height=40mm]{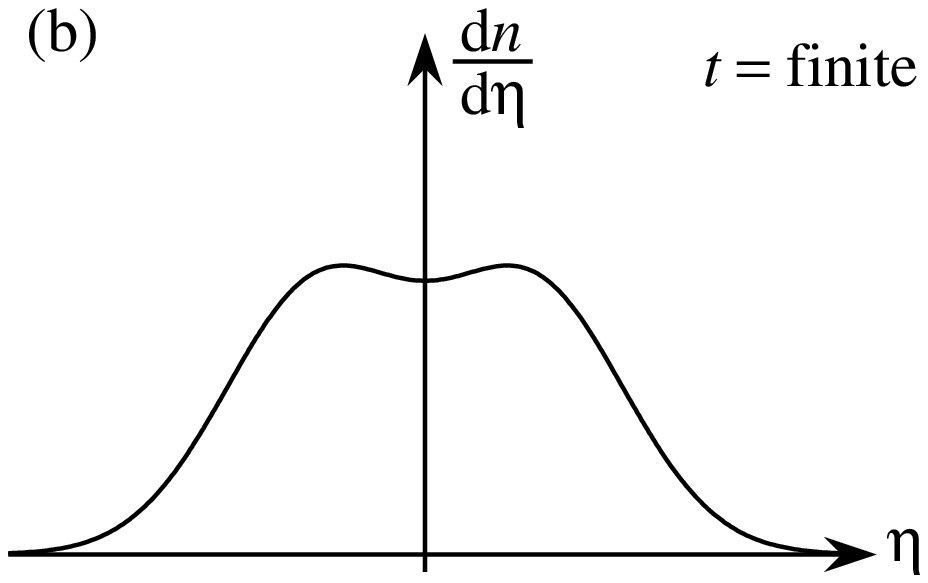}
  \includegraphics[height=45mm]{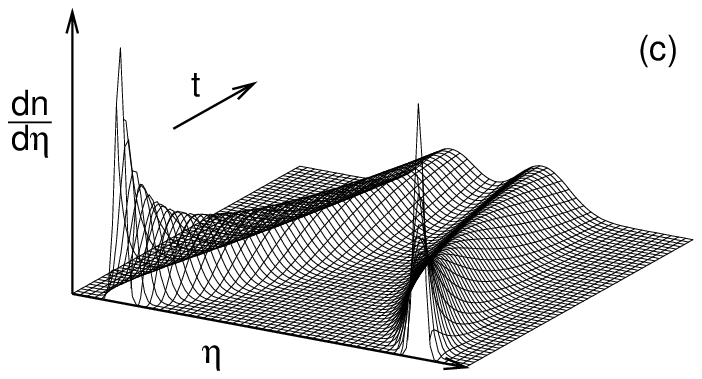}
  \caption{(a) Initial distribution of Eq.~(\ref{intro02}). (b) Final distribution at t = finite. (c) Evolution of Eq.~(\ref{intro02}).}
  \label{fig1}
\end{center}
\end{figure}

The approach named the O-U process is described by the following Fokker-Planck equation,
%
%
\begin{eqnarray}
  \frac{\partial P(y,t)}{\partial t} = \gamma \left[\frac{\partial}{\partial y}y + \frac 12\frac{\sigma^2}{\gamma}\frac{\partial^2}{\partial y^2}\right] P(y,\, t)\:,
\label{intro01}
\end{eqnarray}
where $t$, $\gamma$ and $\sigma^2$ are the evolution parameter, the frictional coefficient and the variance, respectively~\footnote[1]{
The equivalent Langevin stochastic equation with the white noise $f_{\rm w}(t)$ is given as
$$
\frac{dy}{dt} = - \gamma y + f_{\rm w}(t)\:.
$$
\label{foot1}
}. Assuming two sources at $\pm y_{\rm max} = \ln (\sqrt{s_{NN}}/m_N)$
 at $t=0$ and $P(y,\, 0) = 0.5[\delta (y + y_{\rm max})+\delta (y - y_{\rm max})]$,
 we obtain the following distribution function for $dn/d\eta$ (assuming $y \approx \eta$) using the probability density $P(y,\;t)$\cite{Goel:1974,Kampen:1981,Saitou:1980,Hori:1982}
%
%
\begin{eqnarray}
  P(y,\, y_{\rm max},\, t) &=& 
\frac 1{\sqrt{8\pi V^2(t)}}\left\{
\exp\left[-\frac{(y+y_{\rm max}e^{-\gamma t})^2}{2V^2(t)}\right]\right . 
\nonumber\\
 &&\qquad\left .+ \exp\left[-\frac{(y-y_{\rm max}e^{-\gamma t})^2}{2V^2(t)}\right]\, \right\}\, ,
\label{intro02}
\end{eqnarray}
where $V^2(t) = (\sigma^2/2\gamma)p$ with $p = 1-e^{-2\gamma t}$.
The physical picture of Eq.~(\ref{intro02}) with the assumption of $y \approx \eta$ are shown in Fig.~\ref{fig1}. In our approach, it is assumed that $N_{\rm ch}/2$ particles are created at $\pm y_{\rm max}$ at $t=0$. Then these $N_{\rm ch}=(N_{\rm ch}/2+N_{\rm ch}/2)$ particles are evolved according to Eq.~(\ref{intro02}). It is worthwhile to mention that a similar approach for the proton spectra has been given in Ref. \cite{Aggarwal:2000bc}. 
%
%

The contents of the present paper are organized as follows.
In Sec.~II $\eta$ scaling of BRAHMS Collaboration is investigated. In Sec.~III Analyses of $\eta$ distribution by means of Eq.~(\ref{intro02}) are performed. The physical meaning of evolution parameter $\gamma t$ with the frictional coefficient is also considered. In Sec.~IV $z_r=\eta/\eta_{\rm rms}$ ($\eta_{\rm rms}=\sqrt{\langle \eta^2 \rangle}$) scaling is considered. In Sec. V Analysis of $y$ distribution derived $dn/d\eta$ distribution is presented. In the final section concluding remarks are given.
%
\section{Analysis of $\bm{\eta}$ scaling of $\bm{dn/d\eta}$ by BRAHMS Collaboration}
\begin{figure}
\begin{center}
  \includegraphics[height=50mm]{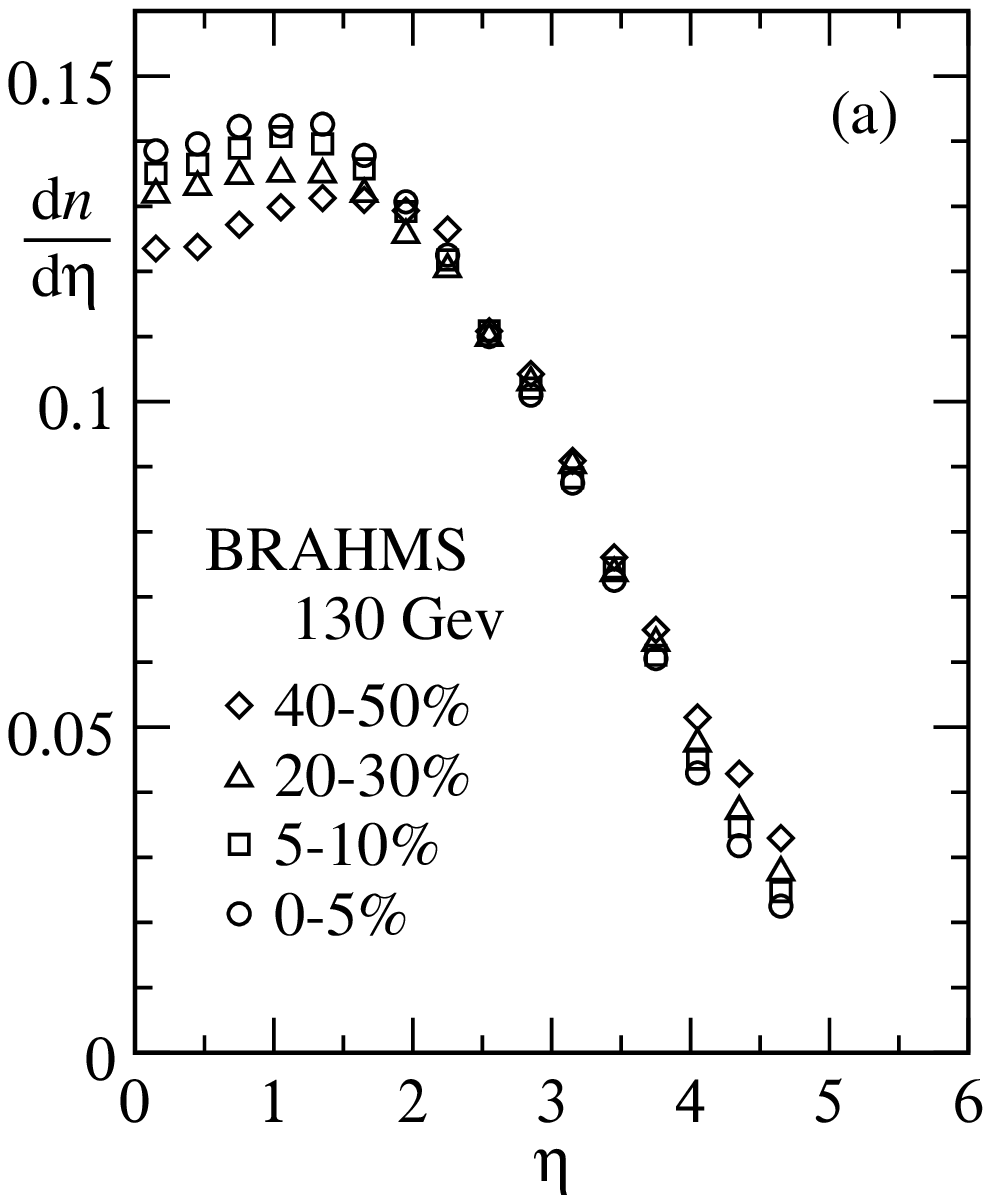}
  \includegraphics[height=50mm]{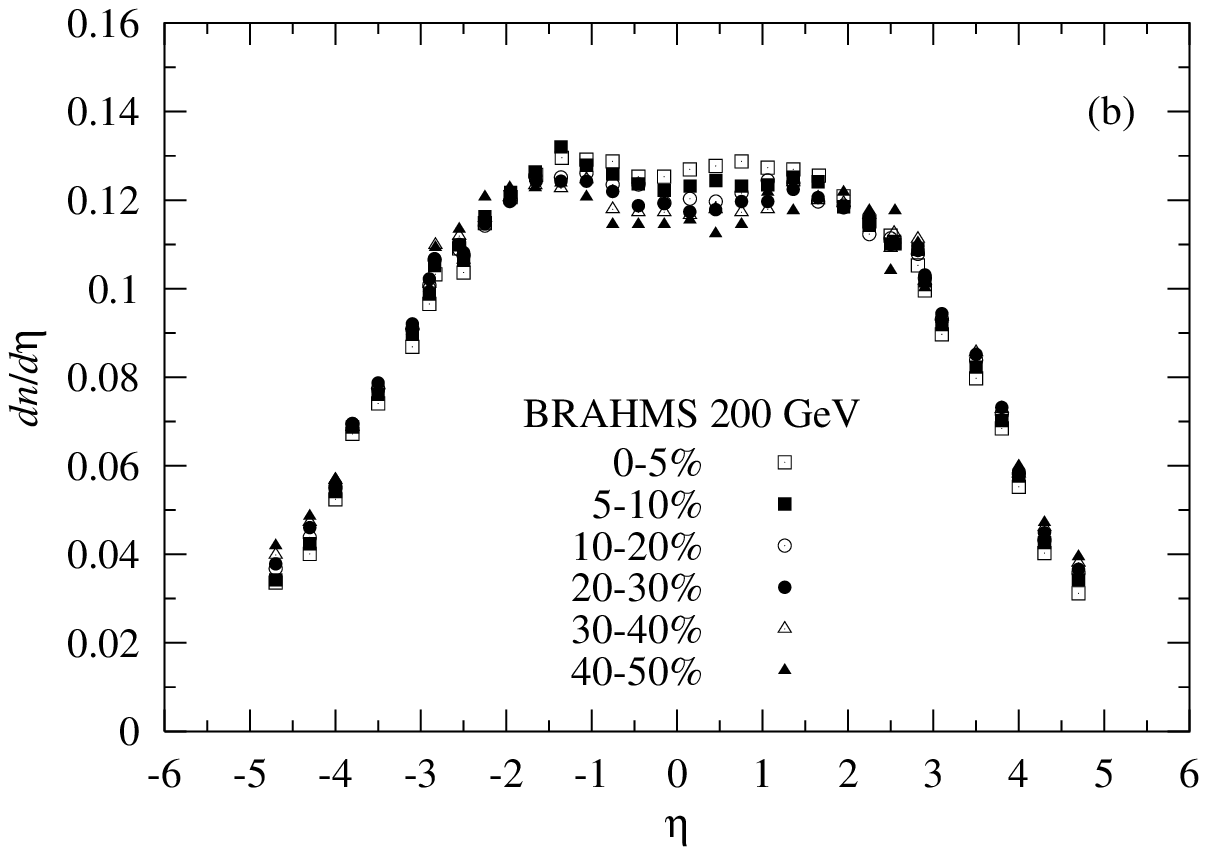}
  \caption{(a) A set of $dn/d\eta$ 's with different centrality cuts at $\sqrt{s_{NN}} = 130$ GeV. Each symbols have error-bars of about $8 \sim 10 \%$ of the magnitude. (b) $dn/d\eta$ with different centrality cuts at $\sqrt{s_{NN}} = 200$ GeV. About the error bars, the situation is the same as (a).}
\label{fig2}
\end{center}
\end{figure}

\begin{figure}
\begin{center}
  \includegraphics[height=50mm]{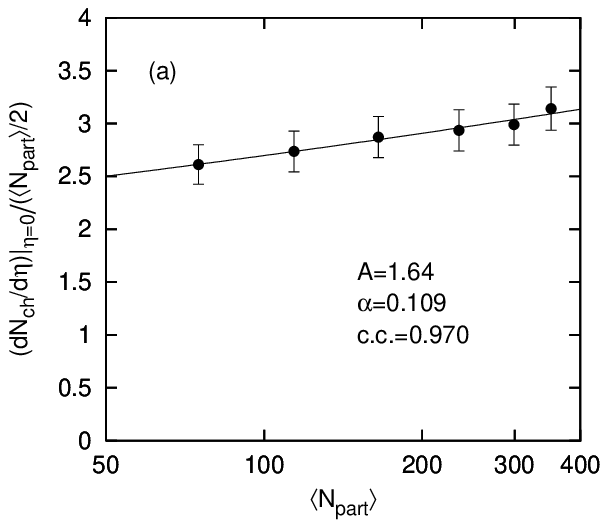}
  \includegraphics[height=50mm]{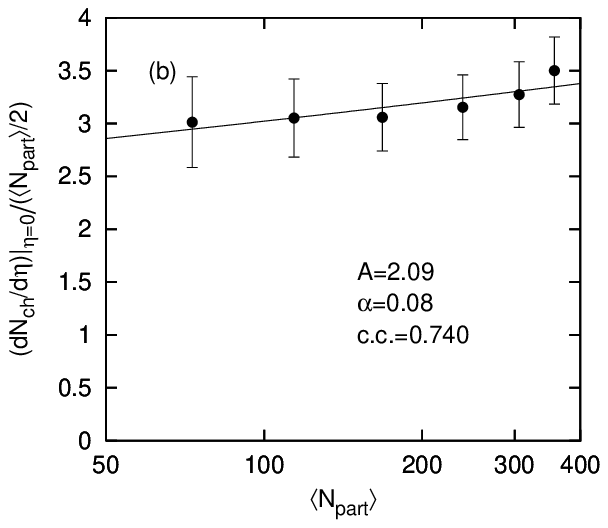}
  \caption{(a) Estimation of parameters A and $\alpha$ at $\sqrt{s_{NN}} = 130$ GeV. The method of linear-regression is used. A$=1.64$, $\alpha=0.109$, and the correlation coefficient (c.c.) is 0.970. A power-like law is seen. (b) $\sqrt{s_{NN}} = 200$ GeV. A$=2.09$, $\alpha=0.08$, and (c.c.)$=0.740$. }
\label{fig3}
\end{center}
\end{figure}

First of all, we consider the problem on $\eta$ scaling in Fig.~\ref{fig2}, plotting the data of $dn/d\eta$ at 130 GeV and 200 GeV. The $\eta$ scaling seems to be held. These distributions show
%
%
\begin{eqnarray}
 \left .\frac{dn}{d\eta}\right |_{\eta = 0} \approx c\ ({\rm constant}).
\label{dndeta01}
\end{eqnarray}
Moreover, we examine the intercept at $\eta=0$. Authors of Ref.~\cite{Aggarwal:2000bc}, WA98 Collaboration, noticed that the intercepts divided by($0.5\langle N_{\rm part}\rangle$) should be described by the power-like law, as 
%
%
\begin{eqnarray}
\label{dndeta02}
\left .(0.5\langle N_{\rm part}\rangle)^{-1} \frac{dN_{\rm ch}}{d\eta}\right |_{\eta = 0} = A\langle N_{\rm part}\rangle^{\alpha}\ ,
\end{eqnarray}
provided that the participants (nuclei) have lost memory and every participant contribute a similar amount of energy to particle production in collisions. Actually it can be said that the power-like law holds, as is seen in Fig~\ref{fig3}. See Tables~\ref{table1} and~\ref{table2}. This physical picture with Eq.~(\ref{dndeta02}) indirectly supports the availability of the stochastic approach. Combining Eqs.~(\ref{dndeta01}) and (\ref{dndeta02}), we have the following relations
%
%
\begin{eqnarray}
c^{\rm Ex} = \frac{1}{N_{\rm ch}} \left .\frac{dN_{\rm ch}}{d\eta}\right |_{\eta = 0},
\label{dndeta03}
\end{eqnarray}
%
%
\begin{eqnarray}
c^{\rm Sp} =\frac{0.5\langle N_{\rm part} \rangle}{N_{\rm ch}} A \langle N_{\rm part} \rangle^\alpha ,
\label{dndeta04}
\end{eqnarray}
where the suffix "Sp" means the semi-phenomenological formula.
Comparisons between Eqs.~(\ref{dndeta03}) and (\ref{dndeta04}) with A and $\alpha$ in Fig.~\ref{fig3} are shown in Tables~\ref{table1} and~\ref{table2}.
%
%
%
%
%
%
\begin{table}
\begin{center}
\caption{Empirical examination of Eqs.~(\ref{dndeta03}) and (\ref{dndeta04}) at $\sqrt{s_{NN}} = 130$ GeV. $\delta c_e = 0.013 \sim 0.015$ and $\delta c_s = 0.010 \sim 0.016$.}
\label{table1}
\begin{tabular}{ccccccc} \hline\hline
centrality (\%) & 40--50 & 30--40 & 20--30 & 10--20 & 5--10 & 0--5  \\ \hline
$\langle N_{\rm part}\rangle$ & 75 & 114 & 165 & 235 & 299 & 352  \\ \hline
$N_{\rm ch}$ & $750\pm 60$ & $1160\pm 90$ & $1720\pm 130$ & $2470\pm 190$ & $3180\pm 250$ & $3860\pm 430$  \\ \hline
$c^{\rm Ex}$ & $0.131\pm \delta c_e$ & $0.134\pm \delta c_e$ & $0.138\pm \delta c_e$ & $0.141\pm \delta c_e $ & $0.143\pm \delta c_e$ & $0.137\pm \delta c_e$\\ \hline
$c^{\rm Sp}$ & $0.131\pm \delta c_s$ & $0.135\pm \delta c_s$ & $0.137\pm \delta c_s$ & $0.141\pm \delta c_s $ & $0.144\pm \delta c_s$ & $0.141\pm \delta c_s$\\ \hline
\end{tabular}
\end{center}
\end{table}

%
%
\begin{table}
\begin{center}
\caption{
The same as Table~\ref{table1} but $\sqrt{s_{NN}}=200$ GeV, $\delta c_e = 0.014 \sim 0.016$ and $\delta c_s = 0.011 \sim 0.016$.}
\label{table2}
\begin{tabular}{cccccccc} \hline\hline
centrality (\%) & 40--50 & 30--40 & 20--30 & 10--20 & 5--10 & 0--5  \\ \hline
$\langle N_{\rm part}\rangle$ & $73\pm 8$ & $114\pm 9$ & $168\pm 9$ & $239\pm 10$ & $306\pm 11$ & $357\pm 8$  \\ \hline
$N_{\rm ch}$ & $890\pm 70$ & $1380\pm 110$ & $2020\pm 160$ & $2920\pm 230$ & $3810\pm 300$ & $4630\pm 370$  \\ \hline
$c^{\rm Ex}$ & $0.124\pm \delta c_e$ & $0.126\pm \delta c_e$ & $0.127\pm \delta c_e$ & $0.131\pm \delta c_e $ & $0.135\pm \delta c_e$ & $0.129\pm \delta c_e$\\ \hline
$c^{\rm Sp}$ & $0.121\pm \delta c_s$ & $0.126\pm \delta c_s$ & $0.131\pm \delta c_s$ & $0.133\pm \delta c_s$ & $0.133\pm \delta c_s$ & $0.129\pm \delta c_s$\\ \hline
\end{tabular}
\end{center}
\end{table}

As is seen in Tables \ref{table1} and \ref{table2}, the intercept at $\eta=0$ is fairly well explained by the semi-phenomenological expression, Eq.~(\ref{dndeta04}). This implies that the stochastic approach may be available, because the participants lost their memory in collision.
%
\section{Analyses of data by Eq.~(\ref{intro02})}
\begin{figure}
\begin{center}
  \includegraphics[height=70mm]{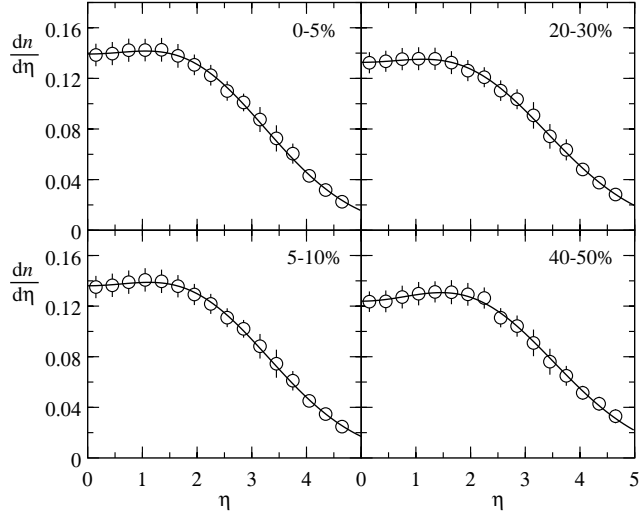}
  \caption{Analyses of $dn/d\eta$ at $\sqrt{s_{NN}} = 130$ GeV by Eq.~(\ref{intro02}). See Table~\ref{table3}.}
\label{fig4}
\end{center}
\end{figure}

\begin{figure}
\begin{center}
  \includegraphics[height=120mm]{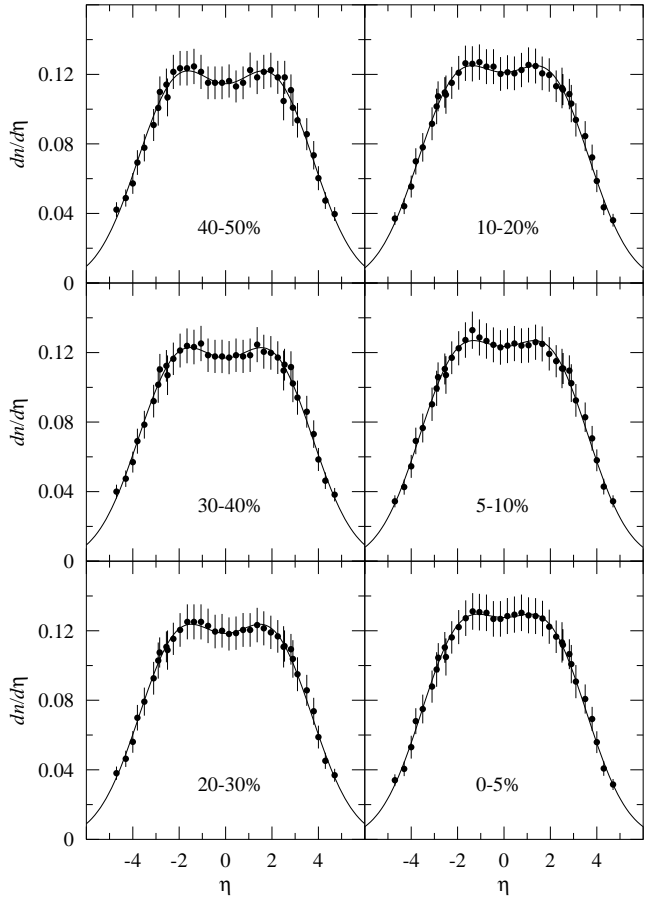}
  \caption{ 
The same as Fig.~\ref{fig4}, but 200 GeV. See Table~\ref{table4}.}
\label{fig5}
\end{center}
\end{figure}

Using the O-U process with two sources, Eq.~(\ref{intro02}), we have analyzed the data. The results at $\sqrt{s_{NN}} = 130$ GeV and 200 GeV are shown in Figs.~\ref{fig4} and~\ref{fig5}, and Tables~\ref{table3} and~\ref{table4}. In our analyses we use Eq.~(\ref{intro02}) the pseudo-rapidity ($\eta$) instead of the rapidity ($y$). As is seen in Tables~\ref{table3} and \ref{table4}, $R=N_{\rm ch}^{\rm (Th)}/N_{\rm ch}$ is always larger than 1. In the measurements of BRAHMS Collaboration, as the observable region is restricted with $|\eta| \le 4.7$, we can conjecture the number of $N_{\rm ch}^{\rm (Th)}$ is always $3\% \sim 7\%$ larger than $N_{\rm ch}$.
%
%
\begin{table}
\begin{center}
\caption{Estimated parameters at $\sqrt{s_{NN}} = 130$ GeV in our analyses by Eq.~(\ref{intro02}) with two sources. Evolution of Eq.~(\ref{intro02}) is stopped at minimum $\chi^2$'s. $\eta_{\rm max} = 4.8$. $R=N_{\rm ch}^{\rm (Th)}/N_{\rm ch}$. $\eta_{\rm rms}=\sqrt{\langle \eta^2 \rangle}$.}\smallskip
\label{table3}
\begin{tabular}{ccccc} \hline\hline
centrality (\%) & 40-50 & 20-30 & 5-10 & 0-5\\ \hline
$N_{\rm ch}^{\rm (Th)}$ & 789$\pm$17 & 1775$\pm$37 & 3273$\pm$68 & 3952$\pm$83\\
$R$ & 1.05 & 1.03 & 1.03 & 1.04\\ \hline
$\eta_{\rm rms} $ & $2.32\pm0.12$ & $2.27\pm0.12$ & $2.24\pm0.12$ & $2.21\pm0.12$\\ \hline
$p$ & 0.841$\pm$ 0.007 & 0.858$\pm$ 0.007 & 0.865$\pm$ 0.007 & 0.871$\pm$ 0.007\\
$V^2(t)$ & 2.79$\pm$0.23 & 2.80$\pm$0.23 & 2.64$\pm$0.21 & 2.56$\pm$0.20\\
$c^{\rm (Th)}$ & 0.124$\pm$0.007 & 0.133$\pm$0.007 & 0.136$\pm$0.008 & 0.139$\pm$0.008\\
$\chi^2/{\rm n.d.f.}$ & 0.877/13 & 0.434/13 & 0.507/13 & 0.758/13\\ \hline
\end{tabular}
\end{center}
\end{table}

%
%
%
%
\begin{table}
\begin{center}
\caption{Estimated parameters at $\sqrt{s_{NN}} = 200$ GeV in our analyses by Eq.~(\ref{intro02}) with two sources. Evolution of Eq.~(\ref{intro02}) is stopped at minimum $\chi^2$'s. $\eta_{\rm max} = 5.4$. $R=N_{\rm ch}^{\rm (Th)}/N_{\rm ch}$, $\eta_{\rm rms}=\sqrt{\langle \eta^2 \rangle}$, $\delta p \approx 0.005$ and $\delta c_t= 0.004 \sim 0.005$.}
\label{table4}
\begin{tabular}{ccccccc} \hline\hline
centrality (\%) & 40-50 & 30-40 & 20-30 & 10-20 & 5-10 & 0-5\\ \hline
$N_{\rm ch}^{\rm (Th)}$& 955$\pm$15 & 1477$\pm$24 & 2158$\pm$34 & 3101$\pm$49 & 4034$\pm$63 & 4881$\pm$76\\
$R$ & 1.07 & 1.07 & 1.07 & 1.06 & 1.06 & 1.05\\ \hline
$\eta_{\rm rms}$ & $2.41 \pm 0.08$ & $2.40 \pm 0.06$ & $2.39 \pm 0.09$ & $2.37\pm0.08$ & $2.35\pm0.08$ & $2.32\pm0.08$\\ \hline
$p$  & 0.854$\pm \delta p$ & 0.859$\pm \delta p$ & 0.862$\pm \delta p$
     & 0.866$\pm \delta p$ & 0.871$\pm \delta p$ & 0.878$\pm \delta p$\\
$V^2(t)$ & 3.169$\pm$0.20 & 3.17$\pm$0.14 & 3.15$\pm$0.19
         & 3.16$\pm$0.19 & 3.10$\pm$0.19 & 3.08$\pm$0.19 \\
$c^{\rm (Th)}$ & 0.115$\pm \delta c_t$& 0.117$\pm \delta c_t$ & 0.118$\pm \delta c_t$ & 0.121$\pm \delta c_t$ & 0.123$\pm \delta c_t$ & 0.128$\pm \delta c_t$\\
$\chi^2/{\rm n.d.f.}$ & 7.2/33 & 5.2/33 & 4.3/33\ & 5.4/33  & 4.9/33 & 5.1/33\\ \hline
\end{tabular}
\end{center}
\end{table}
%
%

The different values of $\chi^2$ in Tables~\ref{table3} and~\ref{table4} are attributed to the magnitude of the error bars at 130 GeV and 200 GeV. 

The intercepts of $dn/d\eta$ at $\eta = 0$ is explained by the following expression in the O-U process,
%
%
\begin{eqnarray}
  c^{\rm (Th)} =  \frac 1{\sqrt{2\pi V^2(t)}}\left\{
\exp\left[-\frac{(\pm \eta_{\rm max}\sqrt{1-p})^2}{2V^2(t)}\right]\, \right\}\, .
\label{analyse01}
\end{eqnarray}
Since our theory is based on the O-U process, the intercept $c^{\rm (Th)}$ is relating to $y_{\rm max}$, the width of $dn/d{\eta}$ and the evolution parameter. \\

Next we consider physical meaning of the evolution parameter $\gamma t$. When we assign the meaning of second [sec] to $t$, $\gamma$ has the dimension of [sec$^{-1}$]. For the magnitude of the interaction region of Au-Au collisions, we assume to be 10 fm. See discussions in Ref.~\cite{Morita:2002av}. See also Tables~\ref{table5} and \ref{table6}. The averaged $\gamma$ [fm$^{-1}$] are almost the same as estimated values from PHOBOS Collaboration \cite{Biyajima:2002wq,Nouicer:2002ks} and ones estimated from the proton spectra at SPS energies in Ref.~\cite{Wolschin:1999jy}.
%
%
\begin{table}
\begin{center}
\caption{Values of $\gamma$ and $\sigma^2$ at $\sqrt{s_{NN}} = 130$ GeV provided that $t\approx 3.3\times 10^{-23}$ sec.}
\label{table5}
\begin{tabular}{cccccc} \hline\hline
centrality (\%) & 40--50 & 20--30 & 5--10 & 0--5 & average\\ \hline
$\gamma$ [fm$^{-1}$] & 0.092 & 0.098 & 0.100 & 0.102 & 0.098 \\
$\sigma^2$ [fm$^{-1}$] & 0.560 & 0.601 & 0.648 & 0.656 & 0.616 \\ 
$\sigma^2/\gamma$ & 6.09 & 6.15 & 6.47 & 6.41 & 6.28 \\ \hline
\end{tabular}
\end{center}
\end{table}
%
%
\begin{table}
\begin{center}
\caption{Values of $\gamma$ and $\sigma^2$ at $\sqrt{s_{NN}} = 200$ GeV provided that $t\approx 3.3\times 10^{-23}$ sec.}
\label{table6}
\begin{tabular}{cccccccc} \hline\hline
centrality (\%) & 40--50 & 30--40 & 20--30 & 10--20 & 5--10 & 0--5 & average\\ \hline
$\gamma$ [fm$^{-1}$] & 0.096 & 0.098 & 0.100 & 0.101 & 0.102 & 0.105 & 0.100\\
$\sigma^2$ [fm$^{-1}$] & 0.714 & 0.723 & 0.724 & 0.733 & 0.729 & 0.738 & 0.727\\
$\sigma^2/\gamma$ & 7.42 & 7.38 & 7.81 & 7.30 & 7.11 & 7.02 & 7.26\\ \hline
\end{tabular}
\end{center}
\end{table}
%
\section{The $\bm{z_r = \eta/\eta}_{\bf rms}$ scaling}
\begin{figure}
\begin{center}
  \includegraphics[height=50mm]{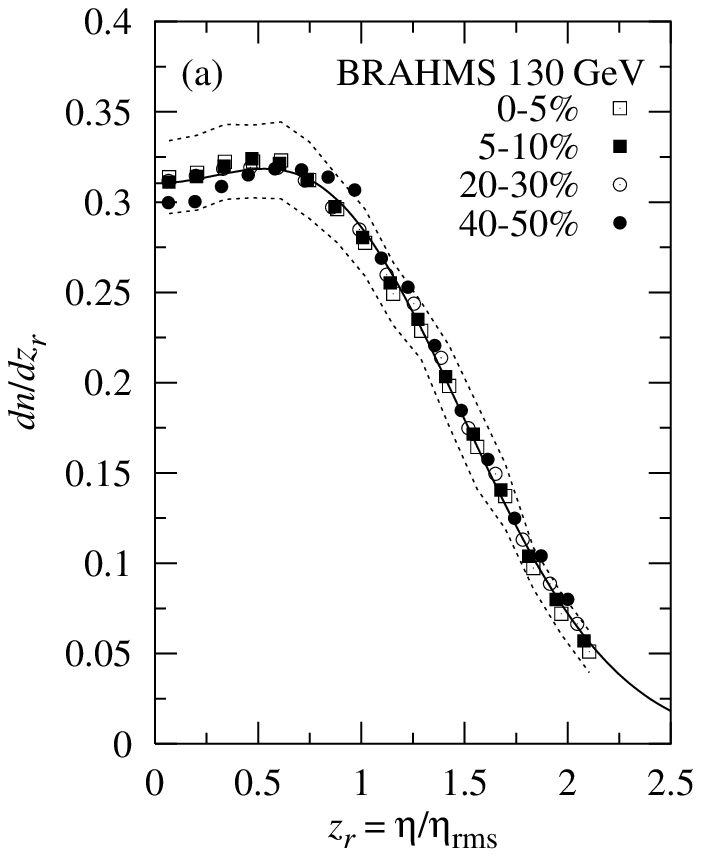}
  \includegraphics[height=50mm]{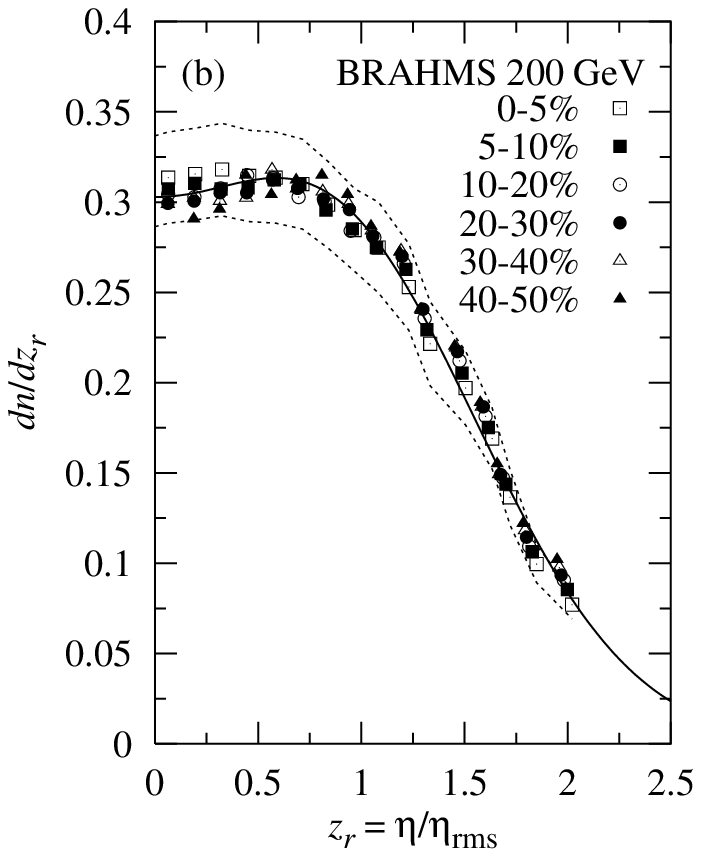}
  \includegraphics[height=60mm]{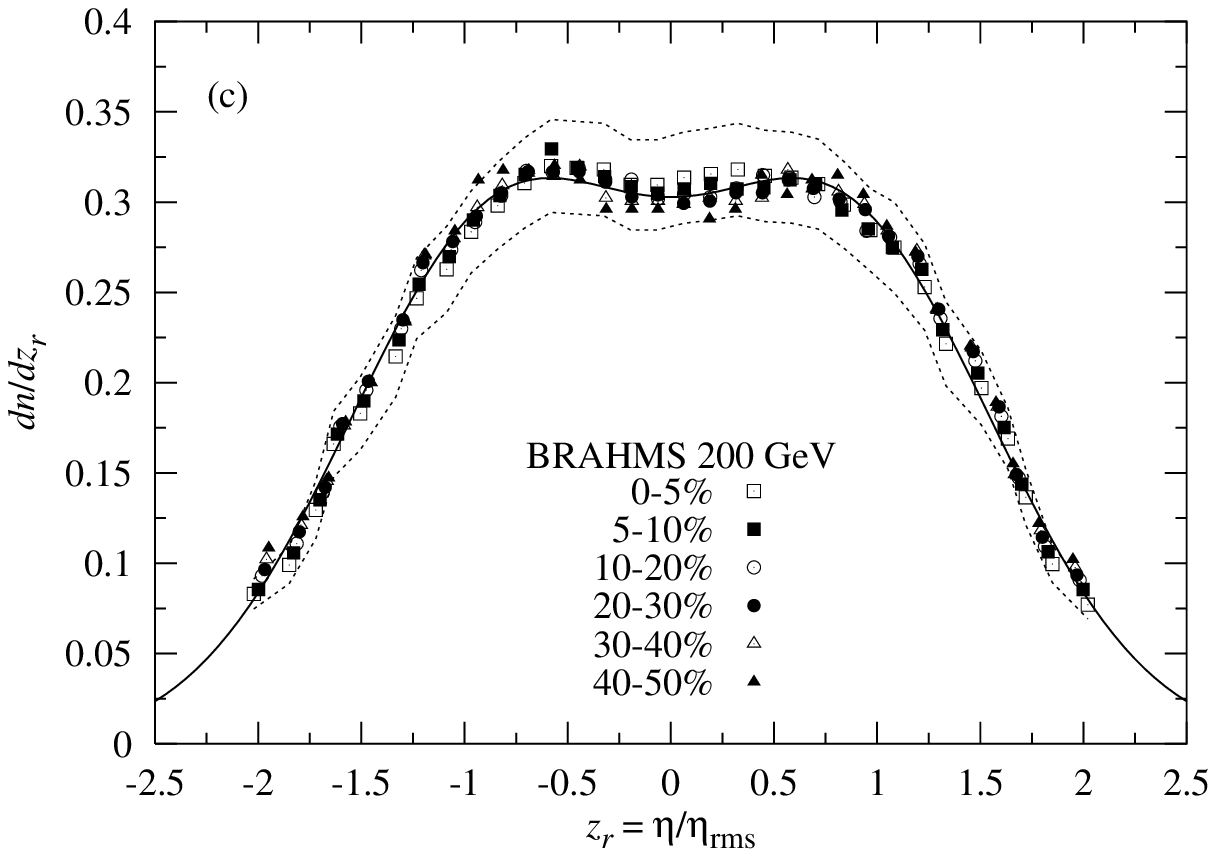}
  \caption{Normalized distribution of $dn/dz_r$ with $z_r=\eta/\eta_{\rm rms}$ scaling and estimated parameters using Eq.~(\ref{zr02}). (a) $\sqrt{s_{NN}} = 130$ GeV, $p = 1-e^{-2\gamma t} = 0.889\pm 0.003$, $V_r^2(t) = 0.527\pm 0.021$ and $\chi^2/{\rm n.d.f.} = 5.4/61$. (b) and (c) $\sqrt{s_{NN}} = 200$ GeV, $p = 1-e^{-2\gamma t} = 0.865\pm 0.002$, $V_r^2(t) = 0.559\pm 0.015$ and $\chi^2/{\rm n.d.f.} = 32.1/189$. (b) is taken from  hemisphere data ($0 \le \eta \le 6$) of Fig.~\ref{fig6}(c). (c) The full space of $dn/dz_r$. The dotted lines represent the magnitude of error-bars in the centrality cut 0-5$\%$.}
\label{fig6}
\end{center}
\end{figure}

To investigate the $z_r=\eta/\eta_{\rm rms}$ scaling which has been proposed in Ref.~\cite{Biyajima:2002at}, we use $\eta_{\rm rms} =\sqrt{\langle \eta^2\rangle} = \sqrt{\sum \eta^2dn/d\eta}$ at $\sqrt{s_{NN}}=130$ GeV and 200 GeV. We can consider the following formula with $z_r$:
%
%
\begin{eqnarray}
\eta_{\rm rms}\frac{d n}{d \eta} = \frac{d n}{d z_r}=f(z_r = \eta / \eta_{\rm rms}).
\label{zr01}
\end{eqnarray}
The right hand side with multiplying $\eta_{\rm rms}$ is obtained from Eq~(\ref{intro02}), as
%
%
\begin{eqnarray}
 \frac{dn}{d z_r} &=& 
\frac 1{\sqrt{8\pi V_r^2(t)}}\left\{
\exp\left[-\frac{(z_r + z_{\rm max}e^{-\gamma t})^2}{2V_r^2(t)}\right]\right . 
\nonumber\\
 &&\qquad\qquad\left .+ \exp\left[-\frac{(z_r - z_{\rm max}e^{-\gamma t})^2}{2V_r^2(t)}\right]\, \right\}\, ,
\label{zr02}
\end{eqnarray}
where $z_{\rm max}=\eta_{\rm max}/\langle \eta_{\rm rms} \rangle$ and $V_r^2(t)=V^2(t)/\eta^2_{\rm rms}$. $\langle \eta_{\rm rms} \rangle$ is the averaged quantity in the set of data. In concrete  analyses of data, $V_r^2(t)$ and $p$ are treated as the free parameters. The $z_r$ scaling at 130 GeV are compared with that of the hemisphere ($0 \le \eta \le 6$) at 200 GeV in Fig.~\ref{fig6} (b). It is difficult to distinguish them without the labels of incident energies.
The behavior of full space is given in Fig.~\ref{fig6} (c). This situation is also observed in analyses of data at 130 GeV and 200 GeV by PHOBOS Collaboration~\cite{Biyajima:2002wq,Nouicer:2002ks}.
%
%
%
\section{Rapidity ($\bm{y}$) distribution derived from $\bm{\eta}$ distribution}
\begin{figure}
\begin{center}
  \includegraphics[height=60mm]{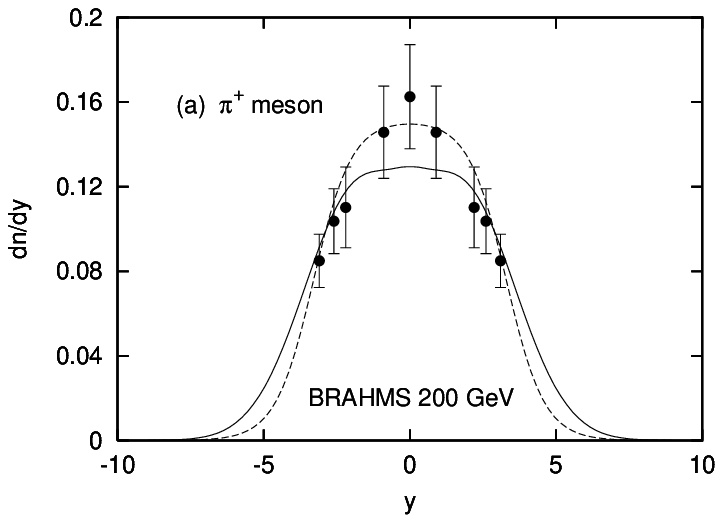}
  \includegraphics[height=60mm]{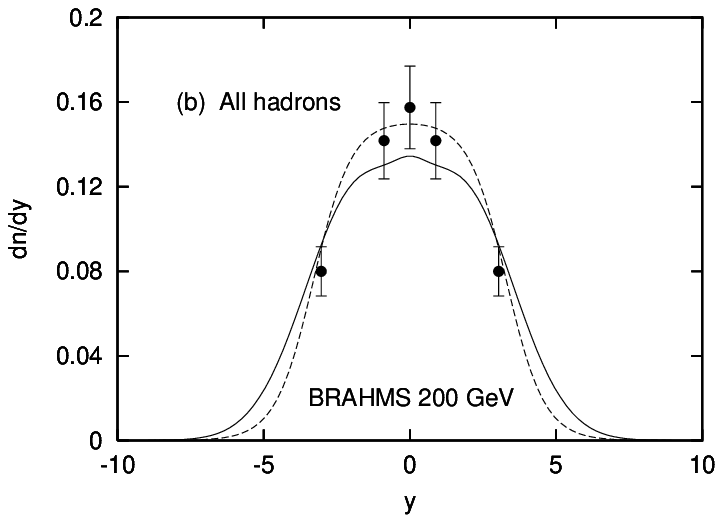}
  \caption{$p$ and $V^2(t)$ are adopted from Tables~\ref{table3} and \ref{table4}. The averaged parameters $p=0.865$ (fixed) and $V^2(t)=3.138$ (fixed) are used. (a) $dn/dy$ of all $\pi$ meson. $m/p_{\rm t}=0.4$ (fixed), $N_{\rm ch}=1503 \pm 77$. (b) $dn/dy$ of all hadrons ($\pi^{\pm}$, $K^{\pm}$, $p$ and $\bar p$). $m/p_{\rm t}=0.5$ (fixed), $N_{\rm ch}=3915 \pm 234$. The dashed lines are obtained from Eq.~(\ref{rapid03}) with $1/c_{\rm N}=0.149$.}
\label{fig7}
\end{center}
\end{figure}

It is well known that one can usually calculate the $\eta$ distribution from the $y$ distribution. In this present study, on the contrary, we consider an inverse problem as follows. First we regard Eq.~(\ref{intro02}) as the correct description of the data, because of small $\chi^2$ values. Using the following formula we can obtain the $y$ distribution~\footnote[2]{
$$
y = \frac 12 \ln \frac{E+p_z}{E-p_z} = \frac 12 \ln\left[\frac{\sqrt{1+  m^2/p_{\rm t}^2+\sinh^2 \eta} + \sinh \eta}{\sqrt{1+ m^2/p_{\rm t}^2+\sinh^2 \eta} - \sinh \eta}\right] = \tanh^{-1} \left(\frac{p_z}E\right) \approx -\ln\tan(\theta/2) \equiv \eta\:.
$$
$$
\eta = \frac 12 \ln \frac{p+p_z}{p-p_z}\quad {\rm and}\quad \frac{dn}{dy} =  \frac{dn}{d\eta} \frac{d \eta}{dy}\:,\ {\rm where}\quad \eta(y)={\rm arcsinh}(\sqrt{M}\sinh y)\:.
$$
For $dn/d\eta=(p/E)dn/dy$, we have $p/E=\cosh \eta/\sqrt{1+ m^2/p_{\rm t}^2+\sinh^2 \eta}$. Moreover, we have confirmed that $\int^{\infty}_{-\infty} (dn/dy) dy=1$ and $\int^{\infty}_{-\infty} (dn/d \eta) d \eta=1$.
\label{foot2}
} as
%
%
\begin{eqnarray}
\label{rapid01}
\frac{dn}{dy} = \frac{\sqrt{M(1+\sinh^2 y)}}{\sqrt{1+M\sinh^2 y}}\frac{d n}{d \eta},
\end{eqnarray}
where $M=1+m^2/p_{\rm t}^2$. The right hand side, $dn/d\eta$, is given as 
%
%
\begin{eqnarray}
\label{rapid02}
 \frac{dn}{d\eta} &=& 
\frac 1{\sqrt{8\pi V^2(t)}}\left\{
\exp\left[-\frac{(\eta(y)+y_{\rm max}e^{-\gamma t})^2}{2V^2(t)}\right]\right . 
\nonumber\\
 &&\qquad\left .+ \exp\left[-\frac{(\eta(y)-y_{\rm max}e^{-\gamma t})^2}{2V^2(t)}\right]\, \right\}\, ,
\end{eqnarray}
where $\eta (y) = {\rm arcsinh}(\sqrt{M}\sinh y)$. From Eq.~(\ref{intro02}) with the averaged parameters $p$ and $V^2(t)$, we obtain $y$ distributions at 200 GeV for $\pi$ meson and all hadrons ($\pi^{\pm}$, $K^{\pm}$, $p$ and $\bar p$). They are compared with the data in Ref.~\cite{Ouerdane:2002gm} in Fig.~\ref{fig7}. The small peak is due to the inverse Jacobian factor. Indeed the data at 200 GeV show these behaviors at $y \approx 0$, even large error bars. To confirm these phenomena, measurements in wider region as well as $y \approx 0$ are necessary.

\begin{figure}
\begin{center}
  \includegraphics[height=80mm]{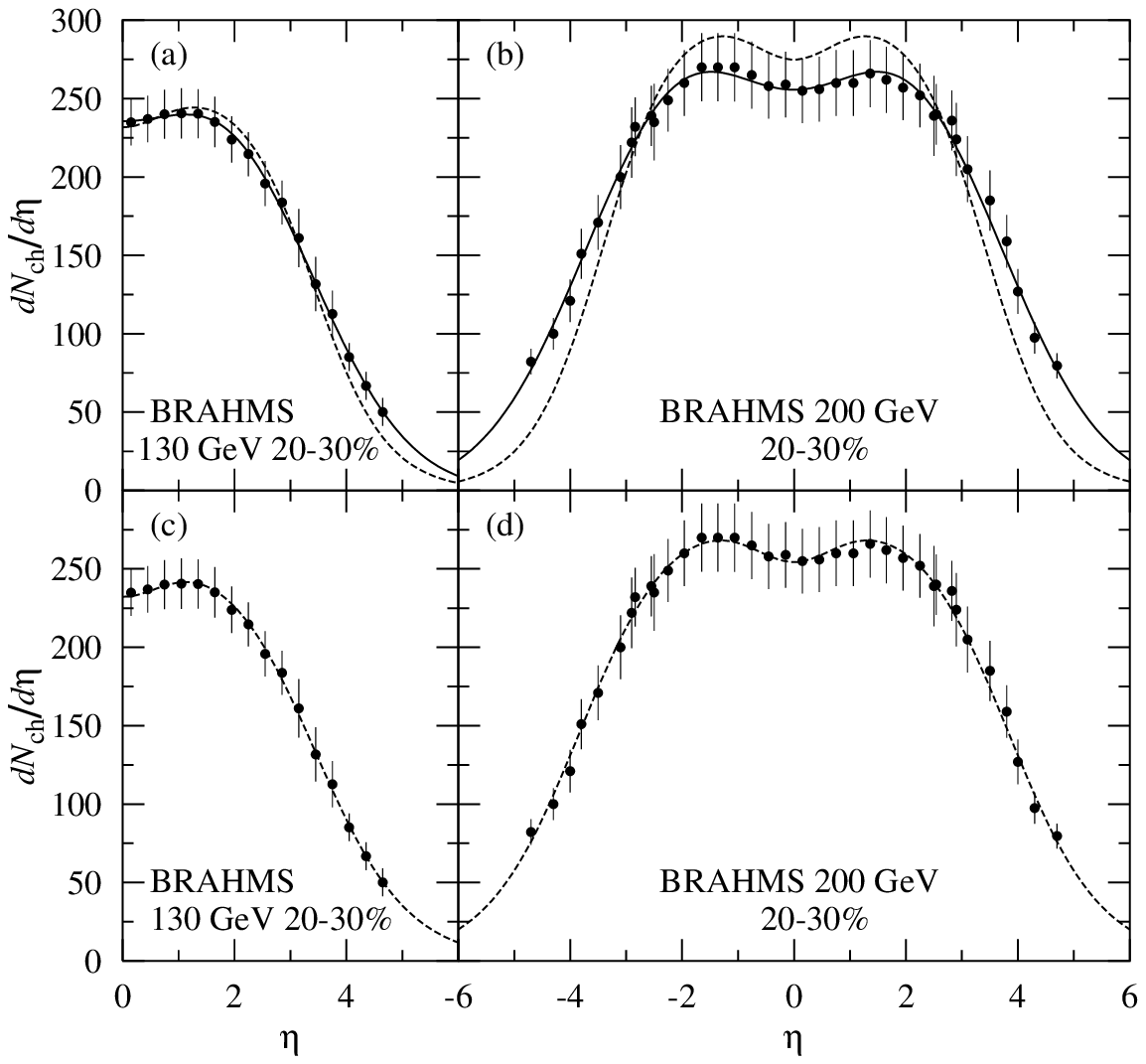}
  \caption{Using Eqs.~(\ref{rapid02}) and (\ref{rapid03}), we calculate $dN_{\rm ch}/d\eta$ (centrality cut 20-30\%) at 130 GeV and 200 GeV. (a) Dashed line is obtained by $y_0 = 3.3$, $d = 0.65$, $N_{\rm ch}^{(\rm Th)} = 1731\pm 34$, $m/p_{\rm t} = 0.5$ and $\chi^2/{\rm n.d.f.} = 12.1/15$. Solid line is obtained by O-U process [$\chi^2/{\rm n.d.f.} = 0.43/13$ from Table~\ref{table3}], (b) Dashed line is obtained by $y_0 = 3.3$, $d = 0.65$, $N_{\rm ch}^{\rm Th} = 2054\pm 31$, $m/p_{\rm t} = 0.5$ and $\chi^2/{\rm n.d.f.} = 134/35$. Solid line is obtained by O-U process [$\chi^2/{\rm n.d.f.} = 4.3/33$ from Table~\ref{table4}]. When $y_0$ and $d$ are treated as free parameters, the following sets of parameters are obtained. (c) $y_0 = 3.32$, $d = 0.83$, $\chi^2/{\rm n.d.f.} = 0.40/13$. (d) $y_0 = 3.72$, $d = 0.83$, $\chi^2/{\rm n.d.f.} = 4.4/33$.}
\label{fig8}
\end{center}
\end{figure}

A phenomenological approach proposed in Ref.~\cite{Eskola:2002qz} (which is named as EKRT) is also shown in Fig~\ref{fig7}.
%
%
\begin{eqnarray}
\label{rapid03}
\frac{d n}{d y}({\rm EKRT})=\frac{1}{c_{\rm N}}\frac{(1+e^{-y_0/d})^2}{(1+e^{(-y-y_0)/d})(1+e^{(y-y_0)/d})},
\end{eqnarray}
where $c_{\rm N}$ is the normalization factor~\footnote[3]{
We have estimated the normalization factor $c_{\rm N}$ as follows
$$
c_{\rm N}=\int^{\infty}_{-\infty} \frac{(1+e^{-y_0/d})^2}{(1+e^{(-y-y_0)/d})(1+e^{(y-y_0)/d})}dy=6.68.
$$
\label{foot3}
}. $y_0=3.3$ and $d=0.65$ are parameters~\footnote[4]{
Notice that a similar expression with its symmetrization can be seen in Ref.~\cite{Ohsawa:1992dj}. A different expression based on the fractional Fokker-Planck equation for $dn/dy$ is found in Ref.~\cite{Rybczynski:2002pn}. Both are proposed for analyses of $pp$ (or $\bar pp$) collisions.
\label{foot4}
} given in Ref.~\cite{Eskola:2002qz}. Eq.~(\ref{rapid03}) also reproduces the both data in Fig.~\ref{fig7}. From Eq.~(\ref{rapid03}) we can calculate $dn/d\eta$ (centrality cut 20-30\%) at 130 GeV and 200 GeV which is presented in Fig.~\ref{fig8}. The coincidences between data and theory are very well, when $y_0$ and $d$ are treated as free parameters.
%
%
%
%
%
\section{Concluding Remarks}
\begin{description}
  \item[1)] We have observed that the behaviors of $\eta$ scaling of $dn/d\eta$ by BRAHMS Collaboration hold fairly well among the various centrality cuts at $\sqrt{s_{NN}} = 130$ GeV and 200 GeV.
  \item[2)] To explain those scaling behaviors, we have assumed that $dn/d\eta$ is governed by the O-U stochastic process with two sources at $\pm y_{\rm max}(\cong \ln \sqrt{s_{NN}}/m_N)$. The intercept of $dn/d\eta$ at $\eta = 0$ is expressed by Eq.~(\ref{analyse01}). See Tables~\ref{table3} and \ref{table4}. The constant $c$'s are reflecting the scaling property relating to the O-U process.
  \item[3)] From the evolution parameter $\gamma t$ and the assumed size of the interaction region of Au+Au collision (10 fm), we have obtained the following value, $\gamma \approx 0.1$ fm$^{-1}$, which is almost the same value as that estimated in Ref.~\cite{Aggarwal:2000bc}.
  \item[4)] From Fig.~\ref{fig6}, it can be said that the $z_r$ scaling holds at 130 GeV and 200 GeV. It is difficult to distinguish them, as compared both data without the labels of incident energies.
  \item[5)] Using Eq.~(\ref{rapid02}) with $\eta$ distributions at 200 GeV, we have calculated the $y$ distributions which explain the data of Ref.~\cite{Ouerdane:2002gm}. The comparison with different approach given in Ref.~\cite{Eskola:2002qz} is also shown. In a future both approaches can be distinguished by the existence of a projection (small peak) at $y \approx 0$.
\end{description}
Finally, it can be concluded that the O-U process is one of possible explanations for the scaling property of $dn/d\eta$ at $\sqrt{s_{NN}} = 130$ GeV and $200$ GeV by BRAHMS Collaboration~\cite{Bearden:2001xw,Bearden:2001qq} as well as distributions by PHOBOS Collaboration~\cite{Nouicer:2002ks}.
%
\section*{Acknowledgements}
One of authors~(M.~B.) would like to thank the Scandinavia-Japan Sasakawa Foundation for financial support, and H.~B\o{}ggild, J.~P.~Bondorf, H.~Ito and K.~Tuominen for their kind hospitality and useful conversations at the Niels Bohr Institute as well as PANIC02.
%


\begin{thebibliography}{99}
\bibitem{Wolschin:1999jy}
G.~Wolschin,
Eur.\ Phys.\ J.\ A {\bf 5}, 85 (1999):
His Fokker-Planck equation is given with a replacing $R(Y,t)$ by $P(y,t)$ as
$$
\frac{\partial P(y,t)}{\partial t}=\frac{1}{\tau_y}\frac{\partial}{\partial y}\left[ (y-y_{eq})P(y,t) \right]+D_y\frac{\partial^2}{\partial y^2}P(y,t) ,
$$
where $y_{eq}$ is relating to the rapidity of the colliding energies.
\end{thebibliography}
\end{document}